\documentclass[pra,aps,superscriptaddress,showpacs]{revtex4}
\usepackage{epsfig}
\usepackage{graphicx}
\usepackage{amsmath}
\usepackage{amscd}

\newcommand{\beq}{\begin{eqnarray}}
\newcommand{\eeq}{\end{eqnarray}}
\def\be{\begin{equation}}
\def\ee{\end{equation}}
\def\ba{\begin{eqnarray}}
\def\ea{\end{eqnarray}}

\begin{document}
\title
{Control of Tripod-Scheme Cold-Atom Wavepackets by Manipulating a
non-Abelian Vector Potential}
\author{Qi Zhang}
\affiliation{Centre of Quantum Technologies and Department of
Physics, National University of Singapore, 117543, Singapore}

\author{Jiangbin Gong}
\affiliation{Department of Physics and Centre for Computational
Science and Engineering, National University of Singapore, 117542,
Singapore}
 \affiliation{NUS Graduate School for Integrative Sciences
and Engineering, Singapore 117597, Republic of Singapore}

\author{C.H. Oh}
\affiliation{Centre of Quantum Technologies and Department of
Physics, National University of Singapore, 117543, Singapore}

\date{\today}
%
\date{\today}
\begin{abstract}
Tripod-scheme cold atoms interacting with laser beams have attracted
considerable interest for their role in synthesizing effective
non-Abelian vector potentials.  Such effective vector potentials can
be exploited to realize an all-optical imprinting of geometric
phases onto matter waves. By working on carefully designed
extensions of our previous work, we show that coherent lattice
structure of cold-atom sub-wavepackets can be formed and that the
non-Abelian Aharonov-Bohm effect can be easily manifested via the
translational motion of cold atoms. We also show that by changing
the frame of reference, effects due to a non-Abelian vector
potential may be connected with a simple dynamical phase effect, and
that under certain conditions it can be understood as an Abelian
geometric phase in a different frame of reference. Results should
help design better schemes for the control of cold-atom matter
waves.
\end{abstract}
 \maketitle

\section{Introduction}

As one of many significant developments in using cold-atoms to
achieve quantum simulations, tripod-scheme cold atoms interacting
with three laser beams have attracted considerable interest for
their role in synthesizing effective non-Abelian vector potentials
\cite{dum,Juzeliunas2004PRL,Juzeliunas2005PRA,Ruseckas2005PRL,monopole,Larson2009PRA}.
Specifically, for cold-atoms with a tripod internal level structure
and in the presence of plane-wave laser fields, a simple
space-independent (as long as the atoms are well inside all the
laser beams) non-Abelian vector potential can be generated. This
result has led to a number of interesting predictions, such as
negative anomalous reflection \cite{Juzeliunas2008PRL}, cold-atom
analog of Datta-Das transistor \cite{Vaishnav2008PRL}, negative
refraction \cite{Juzeliunas2008PRA} and cold-atom analog of the
so-called Zitterbewegung oscillation \cite{Vaishnav2008PRL2,
Merkl2008EPL,qiPRA}.

Recently, we showed that if the laser beams interacting with a
tripod-scheme cold-atom are displaced slowly, then it is possible to
actively imprint a non-Abelian vector potential induced phase
\cite{zee} onto the matter wave \cite{Zhang2009PRA}.  Because the
phases thus obtained are not induced by the translational motion of
the cold atom itself, but by active manipulation of laser-matter
interaction, the imprinting of quantum phases onto matter waves
results in a new type of coupling between internal and translational
motions. Indeed, as shown in Ref. \cite{Zhang2009PRA}, due to the
phase imprinting a wavepacket may develop many interesting
interference patterns and may even be splitted into many
equal-weight copies along a straight line. Because by construction
the phase imprinting due to the laser-beam displacement is
insensitive to the details of the laser beams (such as the laser
intensity, actual speed of the moving laser beam, etc), the active
displacement of laser beams can potentially provide a robust means
of controlling matter wave propagation. Further,  a direct
experimental observation of the phases due to a non-Abelian vector
potential can now be reduced to the observation of the translational
motion of cold atoms.


The purpose of this work is twofold.  First, we consider a few more
scenarios of laser-beam displacement, thus extending and
strengthening our early work \cite{Zhang2009PRA}.  In one scenario
where the laser-beams are moved along a square, we show that it is
possible to split a wavepacket into equal-weight copies on a square
lattice. We further show that it is also possible to form a
triangular lattice of cold-atom sub-wavepackets. In another scenario
where the laser beams are moved along a circle, we show that one may
force a cold-atom wavepacket to ``dance" along a circle of a
different radius or redirect the propagation direction of that
wavepacket.  As seen below, these new scenarios are carefully
designed for our second and more important purpose, namely, to
address two issues of more fundamental interest. The first issue is
on a simple but intriguing demonstration of non-Abelian
Aharonov-Bohm effect via the translational motion of cold atoms. The
second issue is on how our perspectives of quantum phases may change
as we change the frame of reference.  We shall explicate that, by
changing the frame of reference, i.e., by changing from the
laboratory frame to a reference frame moving with the laser beams,
the effect due to an effective non-Abelian vector potential may be
connected with a dynamical phase effect, and under certain
conditions part of this effect may be understood as an Abelian
geometric phase obtained from a different frame of reference. These
results, based on concrete and explicit examples, may enhance our
understanding of the dynamics of tripod-scheme cold atoms and
motivate experiments on the associated control of cold-atom
wavepackets.



This paper is organized as follows. For self-completeness, in Sec.
II we provide some necessary details about the effective non-Abelian
vector potential realized by tripod-scheme atoms. In Sec. III  we
show how moving laser beams along a square can lead to formation of
a square lattice of cold-atom sub-wavepackets. An extension of this
approach can lead to the formation of a triangular lattice of
cold-atom sub-wavepackets. In analyzing the results we show that the
physical effects induced by the non-Abelian vector potential can be
connected to pure dynamical phases in another reference frame. In
the same section we also discuss how such laser manipulation can
serve as a direct approach to the observation of the non-Abelian
Aharonov-Bohm effect. In Sec. IV, we provide an interesting relation
between an evolution matrix as an integral of the non-Abelian vector
potential in the laboratory frame and an Abelian geometric phase in
the frame that moves with the laser beams.  In so doing we consider
the displacement of laser beams along a circle and the resultant
motion of cold-atom wavepackets. Section V concludes this work.

\section{Non-Abelian Vector Potentials Realized by Tripod-Scheme Atoms}
Tripod-scheme atoms refer to four-level atoms interacting with three
laser fields \cite{Bergmann}. We denote the four internal levels as
$|n\rangle$, $n=0-3$. Each of the three transitions $|0\rangle
\leftrightarrow |1\rangle$, $|0\rangle\leftrightarrow |2\rangle$,
and $|0\rangle\leftrightarrow |3\rangle$ is coupled by one laser
field. This coupling scheme can be realized if states $|1\rangle$,
$|2\rangle$, and $|3\rangle$ are degenerate magnetic sub-levels and
the three coupling fields have different polarizations. For
convenience we adopt the same configuration as in Ref.
\cite{Juzeliunas2008PRL}, where two laser beams are
counter-propagating along the $x$-axis and the third laser beam is
along the $z$-axis. The associated internal Hamiltonian under RWA is
given by
\begin{equation}
H_{\text{RWA},4}=\sum_{n=1}^{3}\Omega_{n}|0\rangle\langle n| + h.c.,
\end{equation}
with
\begin{eqnarray}
\Omega_{1}&=&\frac{\Omega_{0}\sin(\xi)}{\sqrt{2}}e^{-ik_{r}^{l}x}, \\
\Omega_{2}&=&\frac{\Omega_{0}\sin(\xi)}{\sqrt{2}}e^{ik_{r}^{l}x},\\
\Omega_{3}&=&\Omega_{0}\cos(\xi)e^{ik_{r}^{l}z},
\end{eqnarray}
where the parameter $\xi$ is set to satisfy $\cos(\xi)=\sqrt{2}-1$,
as in Ref. \cite{Juzeliunas2008PRL}, and $k_{r}^{l}$ is the
wavevector of the laser fields.

The Hamiltonian $H_{\text{RWA},4}$ has two degenerate states with a
null eigenvalue.  We denote these two degenerate states as
$|D_{1(2)}\rangle$ and it is straightforward to find their spatial
dependence as follows \cite{Juzeliunas2008PRL},
\begin{eqnarray}
|D_1\rangle&=&(|\tilde{1}\rangle-|\tilde{2}\rangle)e^{-i\kappa'
z}/\sqrt{2} \nonumber \\
|D_2\rangle&=&\left[\cos(\xi)\left(|\tilde{1}\rangle+|\tilde{2}\rangle\right)/\sqrt{2}
-\sin(\xi)|3\rangle\right]e^{-i\kappa' z},
\end{eqnarray}
where \begin{eqnarray} \kappa'&\equiv& k^{l}_{r}[1-\cos(\xi)],\\
|\tilde{1}\rangle &\equiv & |1\rangle e^{ik^{l}_{r}(x+z)}, \\
|\tilde{2}\rangle &\equiv& |2\rangle e^{-ik^{l}_{r}(x-z)}.
\end{eqnarray}
When the laser field strengths are sufficiently large, $\Omega$ can
be large compared to any possible two-photon detuning induced by
fluctuations in laser frequencies and/or Doppler shift. Then if the
initial state is in the dark-state subspace and if the system
parameters are changing slowly, the internal atomic state can evolve
within the two-dimensional dark-state subspace. As such we focus on
the time evolution in the dark-state subspace only. Clearly then,
the dynamical phase contributed by the internal energy is always
zero because $H_{\text{RWA},4}|D_{1(2)}\rangle=0$. Any state therein
can be spanned as $c_{1}|D_{1}\rangle+c_{2}|D_{2}\rangle$. This
expansion henceforth defines a dark state representation. In this
representation, the mechanical momentum operator becomes
\cite{Juzeliunas2008PRL, groupeqnote}
\begin{eqnarray}
\mathbf{P}^{D}=-i\hbar\tilde{\nabla}+ \hbar \kappa
(\sigma_{x}\hat{e}_{x}+\sigma_{z}\hat{e}_{z}), \label{nonabelianP}
\label{Pgroup}
\end{eqnarray}
where $\sigma_{x,z}$ are Pauli matrices,
$\kappa=\cos(\xi)k_{r}^{l}$, $\tilde{\nabla}$ represents the
gradient in the dark-state representation, and $\hat{e}_{x}$ and
$\hat{e}_{z}$ are the unit vectors along $x$ and $z$. As in Ref.
\cite{Juzeliunas2008PRL}, we also introduce an additional constant
shift to state $|3\rangle$, which accommodates a detuning of the
third laser from resonance by $V_{s}=\hbar
(k_{r}^{l})^2\sin^2(\xi)/2m$ ($m$ is the mass of atom).  Then the
total effective Hamiltonian becomes
\begin{equation} \label{translational}
H^{\text{eff}}_{D}=\frac{(\mathbf{P}^{D})^{2}}{2m}.
\end{equation}
This effective Hamiltonian and the explicit form of $\mathbf{P}^{D}$
in Eq. (\ref{Pgroup}) make it clear that a tripod-scheme atom
interacting with three laser beams effectively synthesize a
non-Abelian vector potential, whose $x$-component is given by $\hbar
\kappa \sigma_x$, $z$-component is given by $\hbar \kappa \sigma_z$,
and $y$-component is zero.

The eigenstates of $H^{\text{eff}}_{D}$ are
\begin{eqnarray} \label{Weigen} \nonumber
|\Psi^{D,\pm}\rangle&=&|g_{\mathbf{k}}^{\pm}\rangle e^{i\mathbf{k}\cdot\mathbf{R}}\\
 &\equiv &\frac{1}{2}\left(\begin{array}{c}
1\mp ie^{i\varphi_{\mathbf{k}}}\\
-i\pm e^{i\varphi_{\mathbf{k}}}
\end{array}
 \right)e^{i\mathbf{k}\cdot\mathbf{R}},
\end{eqnarray}
where $\mathbf{R}$ is the spatial coordinate (for simplicity the
plane-wave normalization factor $(2\pi\hbar)^{3/2}$ is not
included), and $|g_{\mathbf{k}}^{\pm}\rangle $ denotes two-component
internal-state vectors in the dark subspace. Note that
$|g_{\mathbf{k}}^{\pm}\rangle $ depends on the direction of the
wavevector $\mathbf{k}$ via its dependence on
$\varphi_{\mathbf{k}}$, the angle between the $x$-axis and the atom
wavevector $\mathbf{k}$. From Eq. (\ref{nonabelianP}), it is
straightforward to show that the eigenstates $|\Psi^{D,\pm}\rangle$
have momenta $(\mathbf{k}\pm\kappa \hat{\mathbf{k}})\hbar$
respectively, with their energy eigenvalues
$[(k\pm\kappa)\hbar]^{2}/2m$.  Note also that throughout this paper
$\hat{\mathbf{k}}$ represents a unit vector along the vector
$\mathbf{k}$ and $k$ represents the modulus of  $\mathbf{k}$.

\section{Wavepacket Control by Manipulating a non-Abelian Vector Potential}

In Ref. \cite{Zhang2009PRA}, we proposed to control the matter wave
of tripod-scheme atoms by displacing the laser beams along a
straight line. By considering two different frames of reference, in
this section we will first re-analyze our results in Ref.
\cite{Zhang2009PRA}. We then generalize the scheme in
\cite{Zhang2009PRA} by displacing the lasers along a square, as
illustrated in Fig. 1.  As another extension we also consider laser
displacement along an equilateral triangle.

\subsection{Displacing laser-beams along a straight line: two perspectives}

\subsubsection{Perspective from the laboratory frame}
Consider a laser-beam movement along a straight line, e.g., along
the path $A-B$ in Fig. 1.  As what we did in Ref.
\cite{Zhang2009PRA}, it is natural to first analyze this problem in
the laboratory frame. Without loss of generality we assume that the
initial state is given by
\begin{eqnarray} \label{inilab}
|\Psi_{i}^{\text{lab}}\rangle&=&
|g_{\mathbf{k}}(\varphi_{\mathbf{k}}=\pi/2)\rangle e^{i{\bf
k}_0\cdot
{\bf R}} \nonumber \\
&=& \left(\begin{array}{c}
1\\
0
\end{array}
 \right) e^{i{\bf k}_0\cdot {\bf R}} \nonumber \\
 &=& \left(\begin{array}{c}
1\\
0
\end{array}
 \right),
\end{eqnarray}
where we have assumed that the wavevector ${\bf k}_{0}$ associated
with the spatial part of the total wavefunction is zero
\cite{nonzerok}.  Denoting $x=v_{d}t$, where $v_{d}$ is the speed of
the laser beam displacement, then at each spatial point $\mathbf{R}$
in the laboratory frame the evolution of $(c_1,c_2)$ is determined
by
\begin{eqnarray} \label{displacex}
i\frac{d}{d x}\left(\begin{array}{c}
c_{1}\\
c_{2}
\end{array}
 \right)
& =&\left(\begin{array}{cc}
i\langle D_{1}|\frac{\partial}{\partial x}|D_{1}\rangle & i\langle D_{1}|\frac{\partial}{\partial x}|D_{2}\rangle\\
i\langle D_{2}|\frac{\partial}{\partial x}|D_{1}\rangle & i\langle
D_{2}|\frac{\partial}{\partial x}|D_{2}\rangle
\end{array}
 \right)
  \left(\begin{array}{c}
c_{1}\\
c_{2}
\end{array}
 \right) \nonumber \\
& =&-\left(\begin{array}{cc}
0 & \kappa\\
\kappa & 0
\end{array}
 \right)
  \left(\begin{array}{c}
c_{1}\\
c_{2}
\end{array}
 \right) \nonumber \\
 &\equiv & - \hat{G}_{x}\left(\begin{array}{c}
c_{1}\\
c_{2}
\end{array}
 \right),
\end{eqnarray}
where $\hat{G}_x$ is defined as a $2\times 2$ matrix associated with
the displacement along the $x$ direction. This expression is
expected because it reflects the $x$-component of the non-Abelian
vector potential in our dark-state representation (See Eq. (5)).
\begin{figure}[t]
\begin{center}
\vspace*{-0.cm}
\par
\resizebox *{6.5cm}{5.5cm}{\includegraphics*{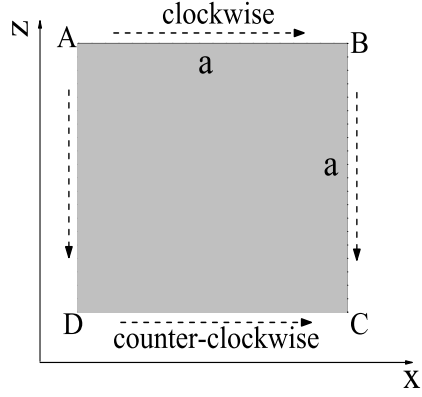}}
\end{center}
\par
\vspace*{-0.5cm}\caption{One scenario of laser beam displacement
along a square for the control of wavepackets of tripod-scheme
cold-atoms. As explained in the text, a cloud of tripod scheme
cold-atoms is simultaneously illuminated by three laser beams. All
the laser beams are moved either clockwise or counter-clockwise from
the starting point $A$, along the square with the length of each
side given by $a$.}
\end{figure}
 One tends to call the evolution matrix determined by the above equation as a non-Abelian geometric phase. However, because here the involved path in the parameter space is not necessarily closed or cyclic, such an evolution matrix is clearly representation-dependent. Nevertheless,
if we move the laser beams such that $e^{ik^{l}_{r}x}=1$, then
because the parameters for both the internal Hamiltonian and the
dark states are subject to cyclic changes, the evolution matrix
associated with Eq. (13) becomes gauge-independent and therefore can
be regarded as a non-Abelian geometric phase. In either case, the
physical effects discussed below are based on the actual solution to
Schr\"{o}dinger's equation and are hence representation-independent.

In addition to this effect due to the non-Abelian vector potential,
a simple dynamical phase also accumulates during the laser
manipulation. In the laboratory frame, the energy of any state
$(c_{1},c_{2})^{T}e^{i\mathbf{k}_{0}\mathbf{R}}=(c_{1}|D_{1}\rangle+c_{2}|D_{2}\rangle)
e^{i\mathbf{k}_{0}\mathbf{R}}$ is contributed by two terms. The
first term is the internal energy of the dark subspace, which is
zero at all times. The second term is the eigenvalues
$[(k_{0}\pm\kappa)\hbar]^{2}/2m$ of the eigenstates given in Eq.
(\ref{Weigen}). In the case of $\mathbf{k}_{0}=0$,  this term is
given by
\begin{equation} \label{constant}
E^{D}_{\text{kin}}=\frac{(-i\hbar\langle
D_{1(2)}|\frac{\partial}{\partial \mathbf{R}}|D_{1(2)}\rangle
)^{2}}{2m}=\frac{\hbar^{2}\kappa^{2}}{2m},
\end{equation}
which is independent of $c_1$ or $c_2$. Hence,  so long as the laser
displacement is sufficiently slow, the internal state will remain in
the dark subspace and the kinetic energy will be necessarily given
by $E^{D}_{\text{kin}}$ for $\mathbf{k}_{0}=0$, which is independent
of $c_1$ or $c_2$.

Based on Eq. (\ref{displacex}) and the dynamical phase determined by
$E^{D}_{\text{kin}}$, one then finds the state evolution during the
passage $A-B$, i.e.,
\begin{eqnarray} \label{Gsplit}
|\Psi(t)\rangle&=&e^{i\hat{G}_{x}v_{d}t}\left(\begin{array}{c} 1\\
0
\end{array}
 \right)e^{i\phi_{d}} \nonumber \\
&=& \frac{1}{2}\left(\begin{array}{c}
1\\
1
\end{array}
 \right)e^{iv_{d}\kappa t}e^{i\phi_{d}}+\frac{1}{2}\left(\begin{array}{c}
1\\
-1
\end{array}
 \right) e^{-iv_{d}\kappa t}e^{i\phi_{d}} \nonumber \\
&=& \frac{1}{2}\left(\begin{array}{c}
1\\
1
\end{array}
 \right)e^{-i\omega_{1} t}e^{i\phi_{d}}+\frac{1}{2}\left(\begin{array}{c}
1\\
-1
\end{array}
 \right) e^{-i\omega_{2}  t}e^{i\phi_{d}},
\end{eqnarray}
where
\begin{eqnarray}
\phi_{d}=-\frac{\hbar\kappa^{2}}{2m}t
\end{eqnarray} is the overall
dynamical phase induced by the kinetic energy term
$E^{D}_{\text{kin}}$, and
\begin{eqnarray}
\omega_{1}&=&-v_{d}\kappa; \nonumber\\ \omega_{2}&=&v_{d}\kappa
\end{eqnarray}
are two frequencies due to the non-Abelian vector potential.
Because, at a time $t$, the total distance of laser displacement in
the $x$ direction is $x=v_{d}t$, the phase factors $\pm v_{d}\kappa
t$ can also be written as $\pm \kappa x$,  depending only on the
distance of laser displacement. This implies the robustness of this
phase acquired by the process, thus manifesting its geometric
nature.  Equation (\ref{Gsplit}) also indicates that the acquired
phases $\pm \kappa a$ at point B are space-independent, and hence it
will maintain ${\bf k}_0=0$ as the wavevector for the spatial part
of the total wavefunction. This further justifies our treatment
here.

Equation (\ref{Gsplit}) for ${\bf k}_0=0$ actually indicates the
existence of two group velocities.  This can be well understood by
finding the derivatives of the frequencies $\omega_{1(2)}$ with
respect to the atomic wavevector $k_{0}$ in the laboratory frame. To
that end we consider a nonzero but small $\delta k_{0}$ in the $x$
direction in the laboratory frame.  We then choose another frame of
reference that moves with atom with a velocity of $\hbar\delta
k_0\hat{e}_{x}/m$. In the new frame of reference the wavevector
becomes zero again,  the laser beams are displaced with a velocity
$\tilde{v}_{d}=v_{d}-\hbar\delta k_{0}/m$, and Eq. (\ref{Gsplit})
can be directly applied with the modified moving speed
$\tilde{v}_d$. Since
\begin{equation} \label{DVK}
\frac{d \tilde{v}_{d}}{dk_{0}}=-\frac{\hbar}{m},
\end{equation}
the two group velocities can then be obtained using Eq. (\ref{DVK})
\begin{eqnarray} \label{groupvgeo} \nonumber
v_{1G}^{\text{lab}}&=&\frac{\partial \omega_{1}}{\partial
k_{0}}=\frac{\partial (-\tilde{v}_{d}\kappa)}{\partial
\tilde{v}_{d}}\cdot\frac{d
\tilde{v}_{d}}{d k_{0}}=\hbar\kappa/m, \\
v_{2G}^{\text{lab}}&=&\frac{\partial \omega_{2}}{\partial
k_{0}}=\frac{\partial (\tilde{v}_{d}\kappa)}{\partial
\tilde{v}_{d}}\cdot\frac{d \tilde{v}_{d}}{d k_{0}}=-\hbar\kappa/m.
\end{eqnarray}
Because these two group velocities associated with the two
components of the evolving state in Eq. (\ref{Gsplit}) are
different, an initial wavepacket of the tripod-scheme cold atoms is
expected to split into two, each of the sub-wavepackets possesses an
internal state $\frac{1}{\sqrt{2}}(1,1)^{T}$ or
$\frac{1}{\sqrt{2}}(1,-1)^{T}$.  This result is nicely confirmed by
numerical simulation results shown in Fig. 2(a). In particular, all
our numerical experiments are based on the full Hamiltonian
$-\frac{\hbar^{2}\nabla^{2}}{2m}+H_{\text{RWA},4}+V_{s}$, and the
initial state is chosen as a Gaussian wave packet instead of a plane
wave.   From Fig. 2(a), it is seen that when the point $B$ is
reached, two sub-wavepackets located at $x=\pm
\frac{\hbar\kappa}{m}t_{B},z=0$ emerge, where $t_{X}$ stands for the
time when point $X$ is reached. Because the length of square shown
in Fig. 1 is given by $a$, we have $t_{B}=a/v_{d}$.

\subsubsection{Perspective from the frame of reference that moves with the laser beams}

In the frame of reference that moves with the laser beams, called
laser frame below, all the three laser beams are static by
construction, but the atom momentum will be changed.  Because the
Hamiltonian in Eq. (\ref{translational}) and its eigenstates in Eq.
(\ref{Weigen}) implicitly assume fixed laser-beams,  this laser
frame can be treated in a straightforward manner. For example, if
the laser beams are moving with a speed $v_{d}$ along the
$x$-direction, then an atom momentum $\mathbf{p}$ in the laboratory
frame will assume a modified momentum
\begin{eqnarray}
\mathbf{p}_{\text{\text{laser}}}=\mathbf{p}-mv_{d}\hat{e}_{x}
\end{eqnarray}
in the laser frame. As a consequence, a wavevector $\mathbf{k}_{0}$
in the laboratory frame will change to
\begin{eqnarray}
\mathbf{k}_{0,\text{laser}}=\mathbf{p}_{L}/\hbar=\mathbf{k}_{0}-mv_{d}\hat{e}_{x}/\hbar
\end{eqnarray} in the laser frame.
For the sake of comparison with our early results, we set
$\mathbf{k}_{0}=0$, and an initial internal state same as that in
Eq. (\ref{inilab}). Thus the total wavefunction in the laser frame
is given by
\begin{equation} \label{inilaser}
|\Psi_{i}^{\text{laser}}\rangle=\left(\begin{array}{c}
1\\
0
\end{array}
 \right)e^{-i\frac{mv_{d}}{\hbar}x}.
\end{equation}

The state in Eq. (\ref{inilaser}) in the laser frame is not an
eigenstate of the Hamiltonian. Nevertheless, it can be written as a
superposition of two eigenstates $|\Psi^{D,+}\rangle$ and
$|\Psi^{D,-}\rangle$ defined previously by Eq. (\ref{Weigen}). Its
time evolution can then be obtained by use of the two energy
eigenvalues associated with $|\Psi^{D,+}\rangle$ and
$|\Psi^{D,-}\rangle$ (with ${\bf k}={\bf k}_{0,\text{laser}}$).
Specifically,
\begin{eqnarray} \label{Dsplit} \nonumber
|\Psi^{\text{laser}}(t)\rangle&=&\frac{1}{2}\left(\begin{array}{c}
1\\
1
\end{array}
 \right)e^{-i\frac{mv_{d}}{\hbar}x}e^{-iE_{+}t/\hbar}\\
 &+&\frac{1}{2}\left(\begin{array}{c}
1\\
-1
\end{array}
 \right)e^{-i\frac{mv_{d}}{\hbar}x}e^{-iE_{-}t/\hbar}.
\end{eqnarray}
where $E_{+}$ and $E_{-}$ are the two eigenvalues $[(k_{0,
\text{laser}}\pm\kappa)\hbar]^{2}/2m$, with $k_{0,
\text{laser}}=-\frac{mv_{d}}{\hbar}$.  One can obtain the same
expression for the wavefunction in the laser frame more formally, by
use of the well-known Galilean transformation for wavefunctions,
namely, $|\Psi(t)\rangle=|\Psi^{\text{laser}}(t)\rangle
e^{i\frac{mv_{d}}{\hbar}x+i\frac{mv_{d}^{2}}{2\hbar}t}$.

Interestingly, the two phase factors in Eq. (\ref{Dsplit}), i.e.,
$e^{-iE_{+}t/\hbar}$ and $e^{-iE_{-}t/\hbar}$, are purely dynamical
phase factors. This is in contrast to our early perspective in the
laboratory frame (where a geometric phase may arise). The two group
velocities corresponding to the two components of the state in Eq.
(\ref{Dsplit}), both along the $x$ direction, can then be obtained
as follows,
\begin{eqnarray} \label{groupv1} \nonumber
v_{+}^{\text{laser}}&=&\frac{\partial E_{+}/\hbar}{\partial
k_{0,\text{laser}}}=-v_{d}+\hbar\kappa/m; \\
v_{-}^{\text{laser}}&=&\frac{\partial E_{-}/\hbar}{\partial
k_{0,\text{laser}}}=-v_{d}-\hbar\kappa/m.
\end{eqnarray}
We can now return to the laboratory frame. Because the laboratory
frame is moving with the laser frame at a velocity
$-v_{d}\hat{e}_{x}$, the two group velocities in the laboratory
frame are also along the $x$ direction, and they are given by
\begin{eqnarray} \label{groupv2} \nonumber
v_{+}^{\text{lab}}&=&v_{+}^{\text{laser}}+v_{d}=\hbar\kappa/m; \\
v_{-}^{\text{lab}}&=&v_{-}^{\text{laser}}+v_{d}=-\hbar\kappa/m.
\end{eqnarray}
This result, which is based entirely on dynamical phase
consideration, is exactly the same as those in Eq. (\ref{groupvgeo})
obtained by considering the integral of the non-Abelian vector
potential in the laboratory frame. Consistent with a previous
finding that a geometric phase may not be Galilean invariant
\cite{SjPLA}, it is still intriguing to see from our concrete
example that the effect of a non-Abelian vector potential (which
leads to a gauge-independent geometric phase for cyclic processes)
may be interpreted totally as that of a dynamical phase in a
different frame of reference. That is, the quantum phases and their
consequences can be traced back to either an integral of the vector
potential in laboratory frame induced by laser beam displacement or
to the dynamical phases in the laser frame (i.e., the time integral
of the energy eigenvalues).  In retrospect, this is not entirely
surprising. The phase due to the vector potential is generated by
active parameter manipulation associated with laser-field
displacement. In a frame of reference in which laser fields are not
moving, there is no such phase induced: observable effects (such as
wavepacket splitting) must be understandable in terms of other
evolution effect, which is purely the dynamical phase here.
\begin{figure}[t]
\begin{center}
\vspace*{-0.cm}
\par
\resizebox *{14cm}{10cm}{\includegraphics*{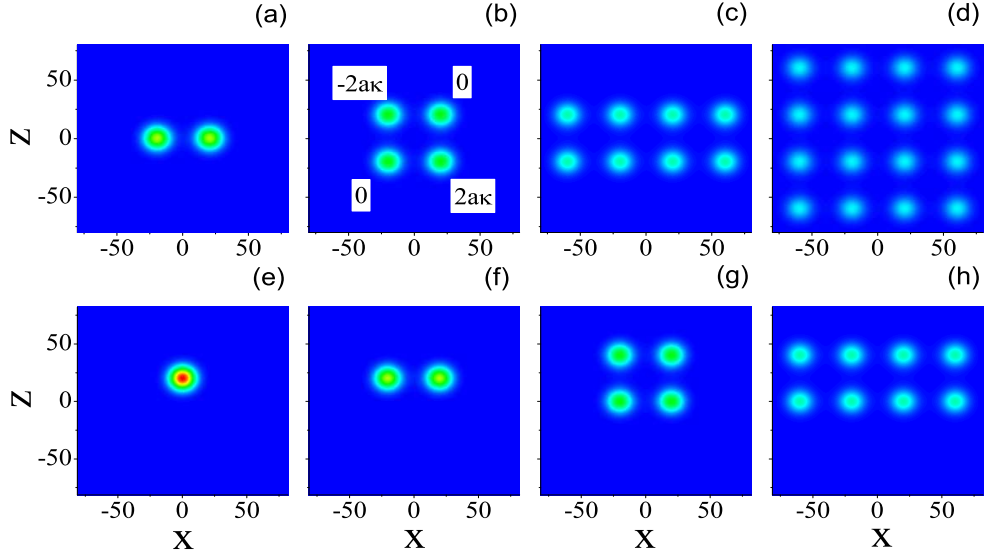}}
\end{center}
\par
 \vspace*{-0.5cm}\caption{(Color Online) Numerical simulation of
the dynamics of tripod-scheme cold-atom wavepackets when all the
three laser beams are slowly displaced along a square shown in Fig.
1. The calculations are based on the full Hamiltonian
$-\frac{\hbar^{2}\nabla^{2}}{2m}+H_{\text{RWA},4}+V_{s}$ (In real
units, we take $m=10^{-25}$ Kg, $\kappa\sim 10^{6}$ m$^{-1}$).
(a)$\rightarrow$(d): wavepacket densities when the points $B$, $C$,
$D$ and $A$ are reached, respectively, in the case of a clockwise
displacement of laser beams. The time intervals for $A-B$ and $B-C$
are $20m/(\hbar \kappa^{2})\sim 0.03s$ and those for $C-D$ and $D-A$
are $40m/(\hbar \kappa^{2})\sim 0.06s$. (e)$\rightarrow$(h):
wavepacket densities in the case of counter-clockwise laser beam
displacement, when the points $D$, $C$, $B$ and $A$ are reached,
respectively. The time intervals for displacement along $A-D$, $D-C$
and $C-B$ are $0.03s$ and that for $D-A$ is $0.06s$. The initial
state is taken as that shown in Eq. (\ref{inilab}). $x$ and $z$ are
in units of $1/\kappa$ and $a=100/\kappa$.}
\end{figure}

\subsection{Formation of a square or triangular lattice of cold-atom sub-wavepackets}

When the laser beams are displaced along the path $A-B$ in Fig. 1,
an initial wavepacket can split into two parts with group velocities
$\pm\hbar\kappa/m$ in the $x$ direction.  Here we first generalize
this scheme by displacing the laser beams along the square shown in
Fig. 1.

The two wavepackets shown in Fig. 2(a) are associated with the two
components shown in the last line of Eq. (\ref{Gsplit}), with the
only difference being that for wavepacket simulations  a Gaussian
profile is imposed on the initial state and hence each component
does not extend to infinity. Because our plane-wave consideration is
seen to describe the actual wavepacket dynamics very well,  we
continue to use plane waves to understand the process along the
square. In addition, because the two wavepackets in Fig. 2(a) are
already separated when the point $B$ is reached, we only need to
treat each term in Eq. (\ref{Gsplit}) separately, rather than
treating the two sub-wavepackets as a whole.

Consider then the second stage of a clockwise laser beam
displacement along the square, which is along the path $B-C$ shown
in Fig. 1, with a velocity $v_{d}$ in the $-z$ direction.  The
effect of the non-Abelian vector potential can be obtained from
\begin{eqnarray} \label{displacez}
i\frac{d}{d z}\left(\begin{array}{c}
c_{1}\\
c_{2}
\end{array}
 \right)
& =&\left(\begin{array}{cc}
i\langle D_{1}|\frac{\partial}{\partial z}|D_{1}\rangle & i\langle D_{1}|\frac{\partial}{\partial z}|D_{2}\rangle\\
i\langle D_{2}|\frac{\partial}{\partial z}|D_{1}\rangle & i\langle
D_{2}|\frac{\partial}{\partial z}|D_{2}\rangle
\end{array}
 \right)
  \left(\begin{array}{c}
c_{1}\\
c_{2}
\end{array}
 \right) \nonumber \\
& =&-\left(\begin{array}{cc}
\kappa & 0\\
0 & -\kappa
\end{array}
 \right)
  \left(\begin{array}{c}
c_{1}\\
c_{2}
\end{array}
 \right) \nonumber \\
 &
 \equiv & -\hat{G}_{z}\left(\begin{array}{c}
c_{1}\\
c_{2}
\end{array}
 \right),
 \label{eqt}
\end{eqnarray}
where a $2\times 2$ matrix $\hat{G}_{z}$ is defined. This effect now
reflects the $z$-component of the non-Abelian vector potential.
Because of the special form of the dark states that have two
wavevectors $k_{r}^{l}$ and $\kappa'$, here the parameter
manipulation is always noncyclic, in the sense that the dark states
do not return to their initial form after the laser displacement.
As such, the evolution matrix as a solution to Eq. (\ref{eqt}) is
necessarily representation-dependent. For this reason, in our fixed
representation for the two dark states $|D_1\rangle$ and
$|D_2\rangle$,  we do not call the evolution matrix determined by
Eq. (\ref{eqt}) as a non-Abelian geometric phase, but only as an
effect due to the non-Abelian vector potential defined above.
(To remove such a representation-dependence one can adopt a unique
gauge proposed in Ref. \cite{cyclic} based on the concept of
parallel frames via re-defining the dark states.  However, it is
found that the resulting non-Abelian gauge potential is too
complicated. Alternatively, we can simply remove the factor
$e^{-i\kappa'z}$ from the dark states defined previously. With the
newly defined dark states, our parameter manipulation will be cyclic
and the obtained solution to  Eq. (\ref{eqt}) will have a desired
gauge-invariance so long as the displacement along $z$ is a multiple
of $2\pi/k_{r}^{l}$.  But most importantly, irrespective of what
representation we use,  the final solution to Schr\"{o}dinger's
equation remains unchanged).

We then proceed by assuming that the initial internal state is
either the first or the second component of Eq. (\ref{Gsplit}).
Apart from an overall phase when the point $B$ is reached, the first
component of Eq. (\ref{Gsplit}) will evolve to
\begin{eqnarray} \label{Gsplitz}
|\Psi_{1,C}(t_C)\rangle&=&\frac{1}{2}e^{-i\hat{G}_{z}v_{d}(t_C-t_B)}\left(\begin{array}{c}
1\\
1
\end{array}
 \right)e^{i\phi_{d}}  \nonumber \\
&=& \frac{1}{2}\left(\begin{array}{c}
1\\
0
\end{array}
 \right)e^{-iv_{d}\kappa (t_C-t_B)}e^{i\phi_{d}}+\frac{1}{2}\left(\begin{array}{c}
0\\
1
\end{array}
 \right) e^{iv_{d}\kappa (t_C-t_B)}e^{i\phi_{d}}
\end{eqnarray}
when point $C$ is reached, where $\phi_{d}$ is again a common
dynamical phase for the internal states $(1,0)^{T}$ and $(0,1)^{T}$.
To seek the group velocities for the two new components in Eq.
(\ref{Gsplitz}), we also need to find the derivatives of the two
frequencies $\pm v_{d}\kappa$ with respect to the wavevector $k_{0}$
in the laboratory frame.  In the same manner as we derive Eqs.
(\ref{DVK}) and (\ref{groupvgeo}), we may consider a $\delta k_{0}$
in the $z$ direction and take advantage of a frame moving at the
velocity $\hbar\delta k_{0}\hat{e}_{z}/m$ relative to the laboratory
frame. Denote again $\tilde{v}_{d}$ as the velocity of the laser
beams in the frame with a zero wavevector for the spatial part of
the total wavefunction, we obtain
\begin{equation} \label{DVKz}
\frac{d \tilde{v}_{d}}{dk_{0}}=\frac{\hbar}{m},
\end{equation}
and the two new group velocities along the $z$ direction,
\begin{eqnarray} \label{groupvgeoz} \nonumber
v_{1G}^{\text{lab}}&=&\frac{\partial (\tilde{v}_{d}\kappa)}{\partial
\tilde{v}_{d}}\cdot\frac{d
\tilde{v}_{d}}{d k_{0}}=\hbar\kappa/m; \\
v_{2G}^{\text{lab}}&=&\frac{\partial
(-\tilde{v}_{d}\kappa)}{\partial \tilde{v}_{d}}\cdot\frac{d
\tilde{v}_{d}}{d k_{0}}=-\hbar\kappa/m.
\end{eqnarray}
These two group velocities correspond to the internal states
$(1,0)^{T}$ and $(0,1)^{T}$, respectively. It can then be predicted
that the wavepacket associated with the first term in the last line
of Eq. (\ref{Gsplit}) will further split into two parts in the $z$
direction, due to the different group velocities $\pm
\hbar\kappa/m$. In the same manner we can predict that the
wavepacket associated with the second term in the last line of Eq.
(\ref{Gsplit}) will also split into two parts. As a consequence,
when the point $C$ is reached, there should be altogether four
sub-wavepackets located at $x=\pm
\frac{\hbar\kappa}{m}\frac{a}{v_{d}},\ z=\pm
\frac{\hbar\kappa}{m}\frac{a}{v_{d}}$. The numerical results shown
in Fig. 2(b) nicely confirm our predictions.

As the laser beams are displaced further along the square shown in
Fig. 1, in principle one can split the wavepacket into $2^{n}$
copies on the $x-z$ plane, forming a beautiful square lattice of
matter waves.  Figure 2(c) and 2(d) show the numerical results of
the distribution of sub-wavepackets when the points $D$ and $A$ are
reached, respectively, in a clockwise order. Remarkably, when the
laser beams are moved back to the initial position $A$, there are
altogether $16$ sub-wavepackets formed.
\begin{figure}[t]
\begin{center}
\vspace*{-0.cm}
\par
\resizebox *{12cm}{10cm}{\includegraphics*{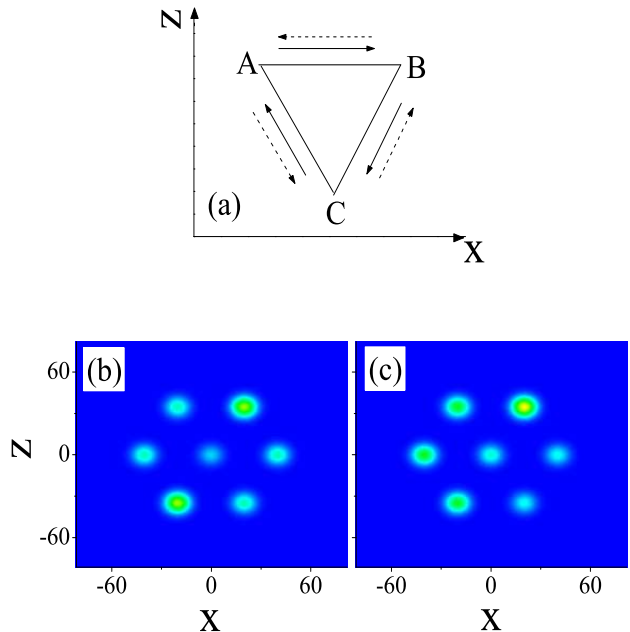}}
\end{center}
\par
\vspace*{-0.5cm}\caption{(Color Online) (a) A schematic plot showing
that the laser-beam displacement is along an equilateral triangle
with the side length $a=100/\kappa$. The laser beams are moved
either clockwise or counter-clockwise from the starting point $A$.
The initial wavepacket is placed at the origin with the internal
state taken the same as that shown in Eq. (\ref{inilab}). (b) The
final distribution of sub-wavepackets, obtained from numerical
simulation of the tripod-scheme cold-atom dynamics when all the
laser beams have been slowly displaced clockwise along the triangle
for one cycle. The calculations are based on the full Hamiltonian
$-\frac{\hbar^{2}\nabla^{2}}{2m}+H_{\text{RWA},4}+V_{s}$ (In real
units, we take $m=10^{-25}$ Kg, $\kappa\sim 10^{6}$ m$^{-1}$). The
time interval to move across each side of the triangle is
$20m/(\hbar \kappa^{2})\sim 0.03s$. $x$ and $z$ are in units of
$1/\kappa$.  (c) Same as in panel (b), but the displacement of the
laser beams is counter-clockwise. }
\end{figure}
It is interesting to examine the relative quantum phases between
each sub-wavepacket. If their relative phases can be easily
calculated and are independent of the details of the manipulation
process, then a coherent lattice of matter wavepackets is created
and they should be useful for atom optics applications such as
atomic interferometry. This is indeed the case here. In particular,
because of a common dynamical phase $\phi_{d}$, the relative phases
between the sub-wavepackets can be easily determined by examining
the imprinted geometric phases, say $\pm v_{d}\kappa t$ shown in
Eqs. (\ref{Gsplit}) and (\ref{Gsplitz}). Take the four
sub-wavepackets in Fig. 2(b) as an example. Other than a common
overall phase,  their quantum phases are found to be $-2a\kappa$,
$0$, $0$, and $2a\kappa$, for the upper-left, upper-right,
lower-left, and lower-right sub-wavepackets, respectively.

It should also be pointed out that, after displacing the laser beams
along a square, a closed path in the parameter space is formed. As
such, the overall effect due to the non-Abelian vector potential
becomes that of  a gauge-independent non-Abelian geometric phase.
Further, because this manipulation is achieved by displacing the
laser beams along different straight lines, our discussions above
also indicate that this geometric phase effect can be equally
understood as arising from the associated dynamical phases in
several laser frames.

The formation of a square lattice of cold-atom sub-wavepackets makes
it clear that it should be equally possible to form other types of
lattices.  A simple consideration shows that if we move the laser
beams along an arbitrary direction, the wavepacket splitting, if it
occurs, must be along the same direction.  As a result, if we move
the laser beams along an equilateral triangle, it should be possible
to form a triangular lattice of cold-atom sub-wavepackets. This is
confirmed by our results shown in Fig. 3.  In particular, Fig. 3(b)
[Fig. 3(c)] depicts the distribution of sub-wavepackets, with
unequal weights, after the laser beams are moved one clockwise
(counter-clockwise) cycle along an equilateral triangle shown in
Fig. 3(a). By repeating this process more sub-wavepackets can be
generated.

\subsection{Non-Abelian Aharonov-Bohm effect}

In Refs. \cite{AB,AB1}, the non-Abelian Aharonov-Bohm effect is
demonstrated by showing different internal states corresponding to
two different paths from a common starting point to a common ending
point, or by showing the non-commutability between different
navigation paths. To make connection between our results and the
context of non-Abelian Aharonov-Bohm effect,  here we consider in
detail what happens to a wavepacket if the laser beams are displaced
counter-clockwise along a square, with the same initial state in Eq.
(\ref{inilab}).  In particular, along the path $A-D$ (see Fig. 1)
the laser beams are displaced in the $-z$ direction. The internal
state will then evolve in accord to Eq. (\ref{displacez}) plus a
dynamical phase $\phi_{d}$. When the point $D$ is reached, then the
system is in the state
\begin{eqnarray}
|\tilde{\Psi}_{D}(t_D)\rangle&=&e^{-i\hat{G}_{z}v_{d}t_D}\left(\begin{array}{c}
1\\
0
\end{array}
 \right)e^{i\phi_{d}} \\ \nonumber
&=&\left(\begin{array}{c}
1\\
0
\end{array}
 \right)e^{-iv_{d}\kappa t_D}e^{i\phi_{d}}.
\end{eqnarray}
Following the same procedure for deriving the group velocities in
Eq. (\ref{groupvgeoz}), we obtain that the phase factor
$-v_{d}\kappa t_{D}=-\kappa a $ will induce one single group
velocity $\hbar\kappa/m$ in the $z$ direction. So when the point $D$
is reached the wavepacket does not split during the passage $A-D$
but moves with a velocity $\hbar\kappa/m$ in the $z$ direction.
 The numerical result shown in Fig.
2(e) confirms this analytical result.

The next stage of a counter-clockwise laser beam displacement is
along the path $D-C$ shown in Fig. 1. Apart from an overall phase,
the total wavefunction when the point $C$ is reached can be obtained
from the phase induced by the non-Abelian vector potential and a
dynamical phase previously defined, i.e.,
\begin{eqnarray}
|\tilde{\Psi}_{C}(t_C)\rangle&=&e^{i\hat{G}_{x}v_{d}(t_C-t_D)}\left(\begin{array}{c}
1\\
0
\end{array}
 \right)e^{i\phi_{d}} \\ \nonumber
&=& \frac{1}{2}\left(\begin{array}{c}
1\\
1
\end{array}
 \right)e^{iv_{d}\kappa (t_C-t_D)}e^{i\phi_{d}}+\frac{1}{2}\left(\begin{array}{c}
1\\
-1
\end{array}
 \right) e^{-iv_{d}\kappa (t_C-t_D)}e^{i\phi_{d}}.
\end{eqnarray}
Associated with the two internal states
$\frac{1}{\sqrt{2}}(1,1)^{T}$ and $\frac{1}{\sqrt{2}}(1,-1)^{T}$, we
find that the two group velocities  are $\pm\hbar\kappa/m$, both in
the $x$ direction.  Thus an initial wavepacket will split into two
parts in the $x$ direction when the point $C$ is reached in a
counter-clockwise order.  Numerical result in Fig. 2(f) again
confirms this.  Further splitting of the wavepackets is shown  in
Fig. 2(g) and 2(h).

By comparing the results associated with clockwise laser beam
displacement and those associated with counter-clockwise
displacement, we can now clearly see a non-Abelian Aharonov-Bohm
effect. In particular, the wavepacket splitting behavior due to
displacement along the path $A-B-C$ (four equal-weight copies in
Fig. 2(b)) is different from that along the path $A-D-C$ (two
equal-weight copies in Fig. 2(f)). This dramatic difference shows
that a non-Abelian Aharonov-Bohm effect can manifest clearly on the
translational motion of cold atoms, rather than on their internal
states considered in Refs. \cite{AB, AB1}.

\section{Relation between a non-Abelian geometric phase and an Abelian geometric phase}

Results in the previous section show that by considering different
frames of reference, new insights into the dynamics induced by laser
beam displacement may be obtained. In this section we extend this
idea and attempt to make a connection between the non-Abelian vector
potential induced phase and an Abelian geometric phase. As a
concrete example, we consider a situation where all the laser beams
are slowly moved along a circle of a radius $r_{L}$. After one
cycle, a closed path in the parameter space is formed, and as such
the integral of the non-Abelian vector potential gives rise to a
non-Abelian geometric phase that is gauge invariant.
\begin{figure}[t]
\begin{center}
\vspace*{-0.cm}
\par
\resizebox *{8cm}{6.5cm}{\includegraphics*{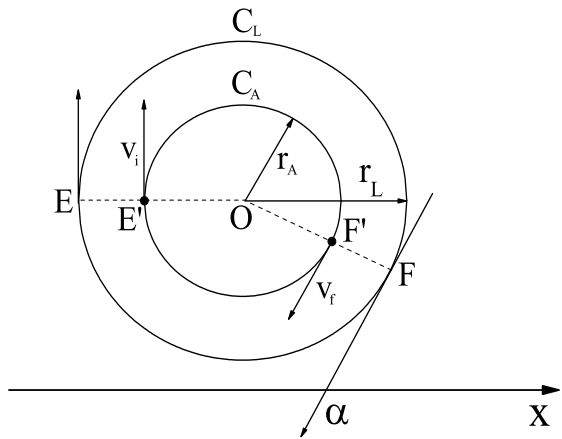}}
\end{center}
\par
\vspace*{-0.5cm} \caption{The scheme of laser beam displacement
along a circle with a radius $r_{L}$, with a starting point $E$. The
atom wavepacket will be forced to dance along a circle with a radius
$r_{A}$ during the process of circular laser beam displacement. The
initial group velocity $\mathbf{v}_{i}$ of a cold-atom wavepacket in
the $z$ direction will be changed to $\mathbf{v}_{f}$ when the point
$F$ is reached.}
\end{figure}

\subsection{Perspective from the laboratory frame}

As shown in Fig. 4, we slowly displace the laser beams from the
starting point $E$ along the circle $C_{L}$ clockwise in the
laboratory frame, with the initial ($t=0$) direction of the laser
displacement in the $z$ direction. The moving speed of the laser
beams is still denoted by $v_{d}$. The initial state in the
laboratory frame is again chosen as that in Eq. (\ref{inilab}).

In a circular laser displacement process, the direction of the laser
displacement always changes, with an angular velocity $v_{d}/r_{L}$.
At time $t$, the laser beams are moving in the direction
$\sin(\frac{v_{d}}{r_{L}}t)\hat{e}_{x}+\cos(\frac{v_{d}}{r_{L}}t)\hat{e}_{z}$.
Let $s=v_{d}t$, at time $t$ we have
\begin{eqnarray} \label{ds}  \nonumber
\frac{dx}{ds}&=&\sin(\frac{v_{d}}{r_{L}}t), \\
\frac{dz}{ds}&=&\cos(\frac{v_{d}}{r_{L}}t).
\end{eqnarray}
Combining Eqs. (\ref{displacex}), (\ref{displacez}) and (\ref{ds}),
one can easily find the evolution of $(c_1,c_2)$  at time $t$
satisfies
\begin{eqnarray} \label{geocir}
i\frac{d}{d t}\left(\begin{array}{c}
c_{1}\\
c_{2}
\end{array}
 \right)&=&-\kappa v_{d}\left(\begin{array}{cc}
\cos(\frac{v_{d}}{r_{L}}t) & \sin(\frac{v_{d}}{r_{L}}t)\\
\sin(\frac{v_{d}}{r_{L}}t) & -\cos(\frac{v_{d}}{r_{L}}t)
\end{array}
 \right)
\left(\begin{array}{c}
c_{1}\\
c_{2}
\end{array}
 \right).
\label{eqt2}
\end{eqnarray}
As explained in the previous section for ${\bf k}_0=0$, the
dynamical phase $\phi_{d}$ given by Eq. (\ref{constant}) is just an
overall phase for any internal state in the dark subspace.
$\phi_{d}$ can hence be neglected.

Interestingly, though Eq. (\ref{geocir}) clearly describes the
effect of a changing non-Abelian vector potential, it however
assumes the very same form as a time-dependent Schr\"odinger
equation of a two-level system. As a result its solution can be
obtained by treating Eq. (\ref{geocir}) as a Schr\"odinger equation
for a time-dependent pseudo Hamiltonian
\begin{equation} \label{pseuH}
H_{p}(t)=-\hbar\kappa v_{d}\left(\begin{array}{cc}
\cos(\frac{v_{d}}{r_{L}}t) & \sin(\frac{v_{d}}{r_{L}}t)\\
\sin(\frac{v_{d}}{r_{L}}t) & -\cos(\frac{v_{d}}{r_{L}}t)
\end{array}
 \right).
\end{equation}
The slow movement of the laser beams induces a slow change of this
pseudo Hamiltonian.  The eigenstates of $H_{p}(t)$ , denoted
$|1(2)\rangle$, are
\begin{eqnarray} \label{Geigen} \nonumber
|1\rangle&=&|g_{\mathbf{k}(t)}^{+}\rangle; \\
|2\rangle&=&|g_{\mathbf{k}(t)}^{-}\rangle,
\end{eqnarray}
where $|g_{\mathbf{k}(t)}^{+(-)}\rangle$ are the internal state
defined in Eq. (\ref{Weigen}), with the angle between the $x$ axis
and $\mathbf{k}(t)$ given by
\begin{eqnarray}
\varphi_{\mathbf{k}(t)}=\frac{\pi}{2}-\frac{v_{d}}{r_{L}}t.
\end{eqnarray}
 The energy eigenvalues of $|1(2)\rangle$ as eigenstates of $H_{p}(t)$  are
\begin{eqnarray} \label{Geigenv} \nonumber
E_{1}&=&-\hbar\kappa v_{d}; \\
E_{2}&=&\hbar\kappa v_{d}.
\end{eqnarray}
Since $E_{1}-E_{2}\neq 0$, the two eigenstates $|1(2)\rangle$ are
necessarily non-degenerate.  Further, because at $t=0$, the initial
state we adopt here (given by Eq. (\ref{inilab})) is exactly the
eigenstate $|1\rangle$ with $\varphi_{\mathbf{k}(t)}=\pi/2$, the
time evolution of this initial state with a slowly changing pseudo
Hamiltonian $H_{p}(t)$ becomes a non-degenerate adiabatic problem.
That is, so long as the change of this pseudo  Hamiltonian is
sufficiently slow, then the adiabatic theorem for a non-degenerate
spectrum guarantees that the time-evolving state will remain on the
instantaneous eigenstate $|g_{\mathbf{k}(t)}^{+}\rangle$ afterwards.

Roughly speaking, a sufficiently slowly changing ``$H_{p}(t)$"
requires \cite{adia}
\begin{equation} \label{adia2}
\frac{|\langle 1(t)|\dot{2}(t)\rangle|}{|E_{1}(t)-E_{2}(t)|}\ll 1.
\end{equation}
Substituting Eqs. (\ref{Geigen}), (\ref{Weigen}) and (\ref{Geigenv})
into Eq. (\ref{adia2}), we find that the above adiabatic condition,
when applied to the pseudo Hamiltonian $H_{p}(t)$,  reduces to
\begin{equation} \label{condition}
r_{L}\gg\frac{1}{\kappa}.
\end{equation}
If this additional condition is satisfied in the laser field
manipulation, then according to the standard adiabatic theorem for a
non-degenerate spectrum, at time $t_{F}$ when an arbitrary point $F$
along the circle is reached, the initial state
$|g_{\mathbf{k}(t=0)}^{+}(\varphi_{\mathbf{k}(t=0)}=\frac{\pi}{2})\rangle=(1,0)^{T}$
will evolve to the instantaneous eigenstate
$|g_{\mathbf{k}(t_{F})}^{+}\rangle$ multiplied by  a geometric phase
factor, namely,
\begin{equation} \label{finplane}
|\Psi_{F}\rangle=|g_{\mathbf{k}(t_{F})}^{+}(\varphi_{\mathbf{k}(t_{F})}=\alpha)\rangle
e^{i\beta}=\frac{1}{2}\left(\begin{array}{c}
1-ie^{i\alpha}\\
-i+e^{i\alpha}
\end{array}
 \right)e^{i\beta},
\end{equation}
where $\alpha$ is the angle between $x$ axis and the tangent line of
the circle at the ending point $F$ shown in Fig. 4. The phase
$\beta$ is given by the following expression that is characteristic
of an Abelian geometric phase,
\begin{equation} \label{Berry}
\beta=\int_{\varphi=\frac{\pi}{2}}^{\varphi=\alpha} \langle
g_{\mathbf{k}}^{+}| \frac{\partial}{\partial \mathbf{k}}
|g_{\mathbf{k}}^{+}\rangle \cdot d\mathbf{k}.
\end{equation}
Because this geometric phase arises from considering a pseudo
Hamiltonian, it could be called as a pseudo Berry-like Abelian
geometric phase. Recalling that the two-component final state given
by Eq. (\ref{finplane}) after moving the laser beams along one cycle
is the result of non-Abelian geometric phases (\ref{geocir})
imprinted onto the initial state $(1,0)^{T}$, all the components
$1-ie^{i\varphi_{\mathbf{k}}}$, $-i+e^{i\varphi_{\mathbf{k}}}$ and
the phase $\beta$ comprise the total effect of the non-Abelian
vector potential. In this sense $\beta$ is just one part of a total
non-Abelian geometric phase.

Figure 4 illustrates what can be predicted from this picture
afforded by the pseudo Hamiltonian defined above.  When $t=0$ the
laser beams are displaced in the $z$ direction and the initial state
is
$|g_{\mathbf{k}(t=0)}^{+}(\varphi_{\mathbf{k}(t=0)}=\frac{\pi}{2})\rangle$;
when the ending point $F$ is reached at $t=t_{F}$, the laser beams
are displaced in the indicated $\alpha$ direction and the final
state is
$|g_{\mathbf{k}(t=t_{F})}^{+}(\varphi_{\mathbf{k}(t=t_{F})}=\alpha)\rangle$.
The group velocity of the time-evolving state can be obtained using
the same method as that used for deriving Eqs. (\ref{groupvgeo}) and
(\ref{groupvgeoz}). It is found that the magnitude of the group
velocity is fixed at $\hbar\kappa/m$, but its direction changes from
$\varphi_{\mathbf{k}(t=0)}=\pi/2$ to
$\varphi_{\mathbf{k}(t=t_{F})}=\alpha$. Thus, the instantaneous
laser displacement direction always coincides with the group
velocity of the wavepacket. Now if the speed of laser displacement
$v_{d}$ is a constant, then one deduces that the atom wavepacket
must dance on a circle $C_{A}$ with a different radius $r_{A}$,
which satisfies
\begin{equation} \label{Rrelation}
\frac{r_{A}}{r_{L}}=\frac{\hbar\kappa/m}{v_{d}}.
\end{equation}
This also suggests that one can easily change the final group
velocity of a wavepacket.  For example, if we need to change the
initial group velocity $\mathbf{v}_{i}$ in the  $z$ direction to the
$\alpha$ direction as shown in Fig. 4, we can displace the laser
beams along the circle $C_{L}$ to point $F$ and then along the
tangent line at point $F$.

All these  predictions have been verified by our numerical
simulations. In particular, in our simulations the initial state is
chosen as a Gaussian wavepacket instead of a plane wave considered
in Eq. (\ref{inilab}). As an example we initially locate the
wavepacket at $x=-50/\kappa,z=0$. The laser beams are displaced with
a velocity $1.5\hbar\kappa/m$ along the circle $C_{L}$ of radius
$r_{L}\approx75/\kappa$. As shown in Figs. 5(a) and 5(c), the
wavepacket indeed dances on a circle of radius
$r_{A}\approx50/\kappa$, satisfying the relation (\ref{Rrelation}).
By contrast, Fig. 5(b) depicts the wavepacket at $t=50m/(\hbar
\kappa^{2})$,  if the laser beams are displaced not along the circle
$C_{L}$ but in the $z$ direction.  The vertical displacement of the
wavepacket indeed shows that the initial group velocity of the
initial state is in the $z$ direction. Figure 5(d) depicts the
wavepacket after the lasers are displaced along the circle $C_{L}$
to point $F$ and then along the tangent line for a duration of
$50m/(\hbar \kappa^{2})$. The result in Fig. 5(d) demonstrates that
the final group velocity is indeed in the direction of the tangent
line at point $F$.
\begin{figure}[t]
\begin{center}
\vspace*{-0.cm}
\par
\resizebox *{10cm}{8.4cm}{\includegraphics*{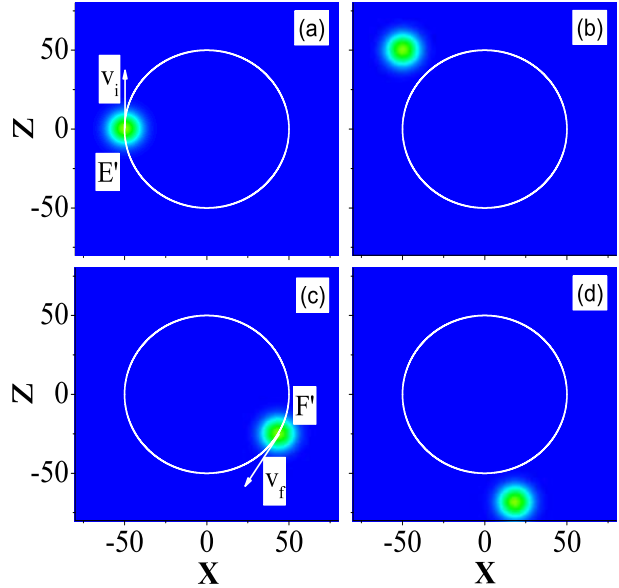}}
\end{center}
\par
\vspace*{-0.5cm} \caption{(Color Online) Numerical simulation of the
wavepacket dynamics of tripod-scheme cold atoms, with the involved
laser beams being slowly displaced along a circle shown in Fig. 4.
The white circle illustrates a possible circular trajectory traced
by the center of the moving wavepacket. $x$ and $z$ are in units of
$1/\kappa$. See the text for details.}
\end{figure}

\subsection{Perspective from a different frame of reference }

In the previous subsection we have argued that one part of a
non-Abelian geometric phase, namely, the $\beta$ phase factor in Eq.
(\ref{Berry}), can be regarded as a pseudo Abelian geometric phase.
In this section, we revisit this connection between a non-Abelian
geometric and an Abelian geometric phase in the laser frame of
reference. As shown below, in this new frame of reference the pseudo
Abelian geometric phase $\beta$ becomes a truly Abelian geometric
phase, i.e., an Abelian geometric phase without introducing a pseudo
Hamiltonian.

As seen before, in the laser frame the wavevector for the spatial
part of the total wavefunction becomes $\mathbf{k}_{0,
\text{laser}}=-\mathbf{v}_{d}m/\hbar$. Note that now the direction
of the laser beam displacement is continuously changing with time.
That is,
\begin{equation}
\mathbf{v}_{d}=v_{d}\left[\cos(\frac{v_{d}}{r_{L}}t)\hat{e}_{z}+\sin(\frac{v_{d}}{r_{L}}t)\hat{e}_{x}\right],
\end{equation}
with its magnitude fixed at $v_{d}$.  Note also that in the laser
frame the wavevector for  the translational part of the total
wavefunction should be given by $\mathbf{k}_{0,\text{laser}}$. One
then finds that in the laser frame the total Hamiltonian in Eq.
(\ref{translational}) can be explicitly written as
\begin{equation} \label{realH}
H_{D}^{\text{eff}}=\frac{\hbar^{2}(k_{0,
\text{laser}}^{2}+\kappa^{2})}{2m}-\hbar\kappa
v_{d}\left(\begin{array}{cc}
\cos(\frac{v_{d}}{r_{L}}t) & \sin(\frac{v_{d}}{r_{L}}t)\\
\sin(\frac{v_{d}}{r_{L}}t) & -\cos(\frac{v_{d}}{r_{L}}t)
\end{array}
 \right).
\end{equation}
The above Hamiltonian assumes the very same form, except the
constant term $\hbar^{2}(k_{0, \text{laser}}^{2}+\kappa^{2})/2m$, as
the pseudo Hamiltonian in Eq. (\ref{pseuH}). Clearly then, the
adiabatic evolution of the eigenstate of $H_{D}^{\text{eff}}$ will
require exactly the same adiabatic condition as Eq.
(\ref{condition}). Under this adiabatic condition, the final result
of the adiabatic process viewed in the laser frame should be given
by the state in Eq. (\ref{finplane}) multiplied by a component due
to $\mathbf{k}_{0,\text{laser}}$, i.e.,
\begin{equation}
|\Psi_{f, \text{laser}}\rangle=\frac{1}{2}\left(\begin{array}{c}
1-ie^{i\alpha}\\
-i+e^{i\alpha}
\end{array}
 \right)e^{i\beta} e^{i\mathbf{k}_{0,\text{laser}}\cdot\mathbf{R}}.
\end{equation}
Returning to the laboratory frame, the wavevector for the spatial
part of the total wavefunction changes from
$\mathbf{k}_{0,\text{laser}}$ to $\mathbf{k}_{0}=0$, so the above
state $|\Psi_{f,\text{laser}}\rangle$ reduces exactly to that
obtained in Eq. (\ref{finplane}).  The two perspectives from
different frames of reference are thus in agreement.  Because here
$H_{D}^{\text{eff}}$ stands for a real Hamiltonian for the
translational motion in the laser frame, the pseudo geometric phase
$\beta$, which is part of the total non-Abelian geometric phase, can
now be understood as a truly Abelian geometric phase in the laser
frame. Thus an interesting relation between non-Abelian and Abelian
geometric phases is established, i.e., in some cases, part of
non-Abelian geometric phase can be interpreted as an Abelian
geometric phase from a different frame of reference.

\section{Concluding Remarks}

To conclude, we have considered two interesting extensions of our
previous study \cite{Zhang2009PRA} of quantum control of matter
waves in tripod-scheme cold-atom systems.  The results have enhanced
our understandings on a number of issues and should motivate further
experimental and theoretical interests in the dynamics of
tripod-scheme cold atoms.

In particular, by displacing the laser beams along a square, it is
shown that the effect of a non-Abelian vector potential can split an
initial wavepacket into $2^{n}$ copies that form a coherent square
lattice. It is also shown that even a triangular lattice of
cold-atom sub-wavepackets can be formed by moving the laser beams
along a triangle. Coherence between the sub-wavepackets thus
generated makes them potentially useful for atom optics
applications. We further discussed how to clearly manifest a
non-Abelian Aharonov-Bohm effect by comparing the wavepacket
profiles obtained by clockwise and counter-clockwise paths along a
square.  Examining  the same process from a different frame of
reference that moves with the laser beams, we find a remarkable
connection between an integral of a non-Abelian vector potential
(which becomes a gauge-independent non-Abelian geometric phase for
cyclic processes) and simple dynamical phase expressions in the
laser frame.

By displacing the laser beams along a circle, we show that a
non-Abelian vector potential effect can force the atom wavepacket to
dance on a circle of a different radius and can also redirect the
propagation direction of that wavepacket. Comparing the perspective
in the laboratory frame and in the frame that moves with the laser
beams, it is shown that one part of the associated non-Abelian
geometric phase in one frame of reference can be understood as an
Abelian geometric phase in another frame of reference.

\section{Acknowledgement}

This work was supported by WBS grant No. R-710-000-008-271 (ZQ and
CH) under the project ``Topological Quantum Computation",  and by
the``YIA" fund (WBS grant No.: R-144-000-195-101) (JG), from the
National University of Singapore.
\bibliographystyle{elsarticle-num}

\begin{thebibliography}{99}
\bibitem{nonzerok}For nonzero ${\bf k}_0$, it can be shown that the
results can approximately apply so long as the laser beam
displacement is much faster than $\hbar k_{0}/m$.
\end{thebibliography}

\end{document}